\documentclass[twocolumn,showpacs,preprintnumbers,amsmath,amssymb,prl]{revtex4}

\usepackage{epsfig,amsmath,amssymb,graphics,color,calc}

\renewcommand{\phi}{\varphi}

\begin{document}

\textbf{Brambilla \textit{et al.} Reply:} van Megen and Williams
(vMW) question~\cite{vMW} our recent claim~\cite{BrambillaPRL2009}
that dense colloidal hard spheres enter at large volume fraction
$\phi$ a dynamical regime not observed in earlier
work~\cite{vMegenPRE1998} and not described by the mode-coupling
theory (MCT) of the glass transition. They claim that our results
are in contradiction to theirs, and suggest that this discrepancy is
due to differences in particle size polydispersity.

\begin{figure} [b]
\psfig{file=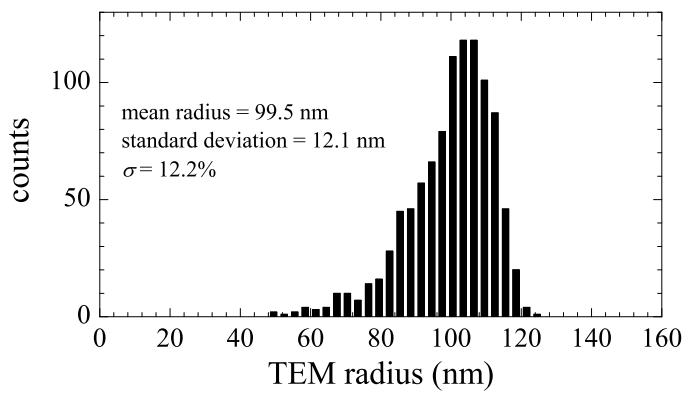,width=9.cm,clip} \caption{\label{fig1} Particle
size distribution as obtained from a sample of 1000 particles imaged
by TEM. The relative polydispersity is $\sigma = 12.2\%$. The
average radius measured by TEM is close to the hydrodynamic
radius measured by DLS, $a \approx 105$ nm. The particle size
reported in Ref.~\cite{BrambillaPRL2009} was somehow higher because
the solvent viscosity had been underestimated.}
\end{figure}

We show in Fig.~\ref{fig1} the particle size distribution obtained
by transmission electron microscopy (TEM). We find $\sigma =
12.2\%$, very close to $\sigma=11.5\%$ as in our
simulations~\cite{ElMasriJSTAT2009}. In this range of $\sigma$, MCT
predicts~\cite{mixture}, and our simulations
reveal~\cite{ElMasriJSTAT2009}, no significant effect due to
differential localization of large and small particles, which thus
cannot account for our data, contrary to vMW's suggestion.

A second explanation suggested by vMW is that a moderate
polydispersity shifts the glass transition to a larger volume
fraction, implying that a non-ergodic sample might become ergodic if
$\sigma$ increases at constant $\phi$. We have considered this
effect. Our simulations~\cite{ElMasriJSTAT2009} show that the effect
is quantitatively modest since, for instance, the position of the
fitted MCT divergence, $\phi_c$, shifts merely by 0.002 when
$\sigma$ changes from 6 to 11.5\%~\cite{ElMasriJSTAT2009}. Taking into account
this $\phi_c$ shift and uncertainties related to volume fraction
determination~\cite{ElMasriJSTAT2009,vMegenPRE1998},
our data are in fact fully consistent with those of
Refs.~\cite{vMegenPRE1994,vMegenPRE1998} up to $\phi \lesssim
\phi_c$.

However, unlike previous work, we have been able to detect ergodic
behavior for samples that have volume fractions above our fitted
$\phi_c \simeq 0.590$, and have discovered that near $\phi_c$ an MCT
description of the data breaks down. Since we have allowed $\phi_c$
to vary to take polydispersity effects into account, our data cannot
be reconciled with MCT in this regime. Indeed, deviations from an
algebraic MCT description can only be cured at the expense of using
unphysical values of the critical parameters. For example, by
imposing $\phi_c = 0.60$ (instead of 0.59 as in
Ref.~\cite{BrambillaPRL2009}), we find that the critical exponent
$\gamma$ in the fitted MCT divergence is as high as $\gamma = 4.5$;
for $\phi_c = 0.605$, the exponent is even higher, $\gamma = 6.8$.
We made similar observations in our two simulated polydisperse hard
sphere models. If these results were solely due to polydispersity,
as claimed by vMW, it should be possible to obtain experimental and
numerical results with less polydisperse samples, say $\sigma
\lesssim 10~\%$, that would cover a range of relaxation times
comparable to that in our work, but still be fully compatible with
MCT. To our knowledge, evidence supporting this scenario is lacking.

Finally, vMW criticize our statement that
this new dynamic regime had not been detected in Ref.~\cite{vMegenPRE1998}
because crystallization intervened.
Indeed, crystallization is not mentioned as an issue
in~\cite{vMegenPRE1998}, although it did intervene in
\cite{vMegenPRE1994}, where a sample with
$\sigma \approx 4~\%$
was studied. vMW emphasize that the samples
with $\phi > \phi_c$ in Ref.~\cite{vMegenPRE1998}
are not ergodic, even if a larger time window is used: they mention
a more recent work~\cite{vMegenPRL2008} where the non-ergodic
aging dynamics of a sample with $\phi=\phi_c + 0.01$ is studied over 5 days.
From the fit of the relaxation time $\tau_\alpha(\phi)$ discussed
in~\cite{BrambillaPRL2009}, we estimate that $\tau_{\alpha}$
grows by a factor $\sim 500$ when $\phi$ increases from $\phi_c$ to
$\phi_c+0.01$. Assuming a similar behavior for the sample studied
in~\cite{vMegenPRL2008}, no equilibration is to be expected
before several hundreds of days,
much longer than the largest waiting time
in that work.
Similar arguments apply to the sample at $\phi = 0.583$
 in~\cite{vMegenPRE1998}.


To conclude, our data show no discrepancy with earlier work, but
explore a broader dynamical range, including an activated regime
that has not been accessed before.

G. Brambilla, D. El Masri, M. Pierno, L. Berthier, and L. Cipelletti
\\ {\small LCVN UMR 5587, Universit\'e Montpellier 2 and CNRS,
34095 Montpellier, France}
\newline
G. Petekidis
\\ {\small IESL-FORTH and Department of Material Science and Technology,
University of Crete, 71110 Heraklion, Greece}
\newline
A. Schofield
\\ {\small School of Physics, University of Edinburgh, Mayfield Road,
Edinburgh EH9 3JZ, UK}

\end{document}